\documentclass[twoside]{mhd}
\usepackage{amsmath}
\usepackage{graphicx}
\usepackage{caption, color}
\usepackage{subcaption}
\usepackage{multirow}
\usepackage{float}
\usepackage{cite}
\usepackage{bm}
\usepackage{textcomp}
\mhdhead{40}{1}{1}

\newcommand{\new}[1]{\textcolor{red}{#1}}

\title{Transient Behaviour of Electro-Vortex Flow in a Cylindrical Container}

\author{K.~Liu\inst{1,2*}, F.~Stefani\inst{2}, N.~Weber\inst{2}, T.~Weier\inst{2}, B.W.~Li\inst{1}}

\institute{School of Energy and Power Engineering, Dalian University of Technology, Dalian, 116024, China 
	\and Helmholtz-Zentrum Dresden-Rossendorf, Bautzner Landstra{\ss}e 400, 01328 Dresden, Germany
}

\begin{document}
	\maketitle
	
	\begin{abstract}
		This study is a continuation of a combined experimental and numerical investigation \cite{Ke2020} of the flow of the eutectic alloy GaInSn inside a cylindrical vessel exposed to a constant electrical current. The emerging electro-vortex flow (EVF), caused by the interaction of the current, which is applied through a tapered electrode, with its own magnetic field might have both detrimental and advantageous effects in liquid metal batteries (LMBs). While the former work \cite{Ke2020} was mainly concerned with time-averaged results, this paper focuses on the transient behaviour of the EVF which becomes most relevant under the influence of an external axial field. The additional Lorentz force, generated by the interaction of the imposed current with the vertical component of the geomagnetic field ($b_z$), drives the ordinary EVF jet flow into a swirling motion. The velocity distributions and motion characteristics, such as spiral streamlines, and shortened and irregularly swinging jet regions, are investigated. The mechanism is analysed in detail for $b_z=-25.5\hspace{1.5pt}\mu$T. The maximum angular velocity of the rotating jet is basically linearly dependent on $b_z$, at least for the values studied here. A good agreement between the transient simulation and experimental result is shown.

		\textbf{Key words}: electro-vortex flow; liquid metal battery; external magnetic field; jet flow; transient behaviour
	\end{abstract}
	
	\section{Introduction}	
		Quite generally, electro-vortex flows (EVF) emerge when currents converge or diverge inside a liquid conductor \cite{Bojarevi1998,Shercliff1970}. EVFs are not classified as flow instabilities since they do not need any critical threshold to set in. In liquid metal batteries (LMBs), EVFs usually appear close to the top and bottom due to the differences of the diameters of the current collectors and the connected cables \cite{Weber2015,Kelley2018}. In the cathodes of LMBs, some moderate EVF is considered favourable since the enhanced mass transfer of the liquid alloy \cite{Kelley2018,Ashour2018,Weber2018,Weier2017,Weber2020} could suppress the formation of solid intermetallic compounds, which otherwise might arch or stick on the inner wall of the LMBs which eventually results in a short circuit. On the other hand, EVFs in LMBs should not be too strong either, otherwise a short circuit might occur due to the flow triggered deformation of the electrolyte layer \cite{Stefani2016,Herreman2019}. Likewise, those deformations might also result from flow instabilities such as the Tayler instability, interfacial instabilities, and the Rayleigh-Bénard instability. The potential presence of all these instabilities, and their possible interaction with each other as well as with the EVF, has transformed LMBs into quite sophisticated (magneto-)hydrodynamic systems, whose full understanding needs more fundamental investigations.

\begin{figure}[htbp]
			\centering
			\includegraphics[width=\textwidth]{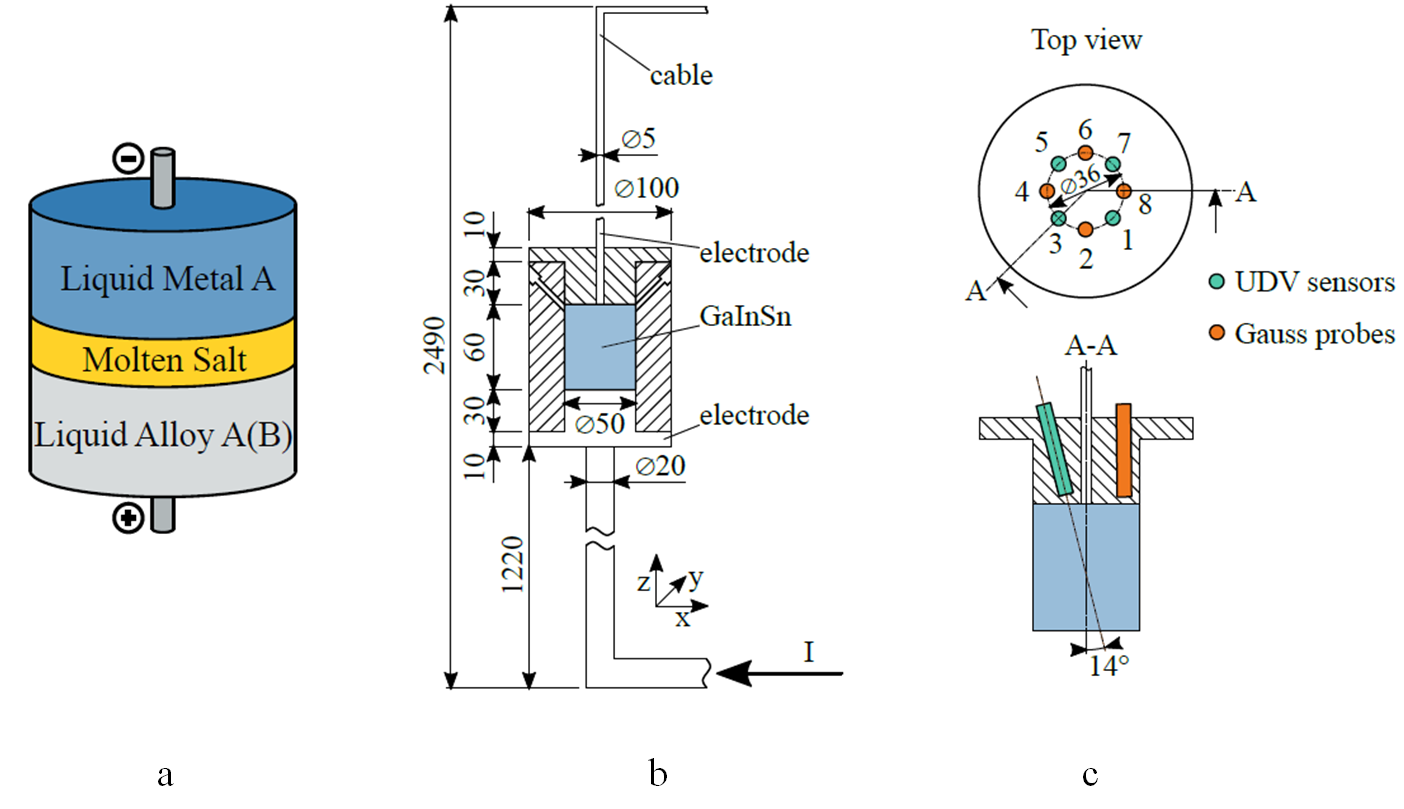}
			\caption{A schematic of an LMB (a), and the EVF experiment: (b) Sketch of the experiment. (c) Partially enlarged $(1.5:1)$ cross-sectional view showing the distribution of the UDV sensors and the holes for the magnetic field measurements in more detail. }
			\label{fig:experimental_set-up}
		\end{figure}

		Continuing our previous work \cite{Ke2020}, the present paper is devoted to a more detailed investigation of the time-dependent characteristics of the EVF in a cylindrical container, filled with the GaInSn liquid metal alloy. As explained in \cite{Ke2020}, the fluid region in our experimental setup (Figure \ref{fig:experimental_set-up}\hspace{1pt}b) could be considered as one of the layers of a liquid metal battery \cite{Weber2015} (as shown in Figure \ref{fig:experimental_set-up}\hspace{1pt}a). The replacement of those liquid metals as typically used in LMBs \cite{Lalau2016,Ouchi2014,Kim2013,Shin2015,Lu2014,Wang2014,Bradwell2012}, by the alloy GaInSn ($67:20.5:12.5 wt\%$) \cite{Dobosz2018} is mainly motivated by its easy handling related with its low melting temperature of about 10\hspace{1.5pt}\textcelsius.

		In our previous work \cite{Ke2020}, the physical and mathematical models, together with boundary and initial conditions were described and the numerical solver was introduced. Moreover, we validated the grid independence, illustrated the distributions of electric current density and Lorentz force, and discussed the influences of the geomagnetic field and Joule heating on the distribution of time-averaged velocity field. As observed in \cite{Ke2020}, the vertical component of the geomagnetic field ($b_z$) plays an important role in affecting the EVF’s flow field structure. To better understand the role of $b_z$, in this paper the flow field is investigated by analysing the spatial distribution and the helicity of the transient velocity field, the variations of the time-averaged velocity streamlines and the maximum angular velocity at $z=58$\hspace{1.5pt}mm for different $b_z$ values, as well as the time-dependent 3D flow field structures for the specific value $b_z=-25.5\hspace{1.5pt}\mu$T. The helical flow structures and motion characteristics are illustrated, and the formation mechanism is explained in detail. Finally, in order to validate our time-dependent numerical simulations, the temporal evolution of the numerical velocity is compared with measurement data, which also goes beyond the comparisons of the time-averaged velocities of the numerical and experimental results as carried out in our former article \cite{Ke2020}. The paper concludes with a summary and an outlook on further research.
	
	\section{Physical model and mathematical model}
	
		For the sake of clear understanding, we introduce briefly our physical and mathematical models first. As shown in Figure \ref{fig:experimental_set-up}\hspace{1pt}b, our setup mainly includes five conductive parts: the top and bottom cables, the top and bottom electrodes, and the GaInSn liquid metal. A current (80\hspace{1.5pt}Ampere) flows in from the bottom cable, and out from the top cable. For simulating the EVF correctly, the energy equation, the momentum equation, the continuity equation, the Poisson-type equations for the pressure and the electric potential, and the Biot-Savart law for the magnetic field are solved simultaneously. A more detailed descriptions of the physical model, the mathematical model, the initial and boundary conditions, and the numerical implementation can be found in \cite{Ke2020}, and will not be reiterated here.

	\section{Results and discussions}
	\subsection{Velocity distributions}
\begin{figure}[htbp]
			\centering
			\includegraphics[width=\textwidth]{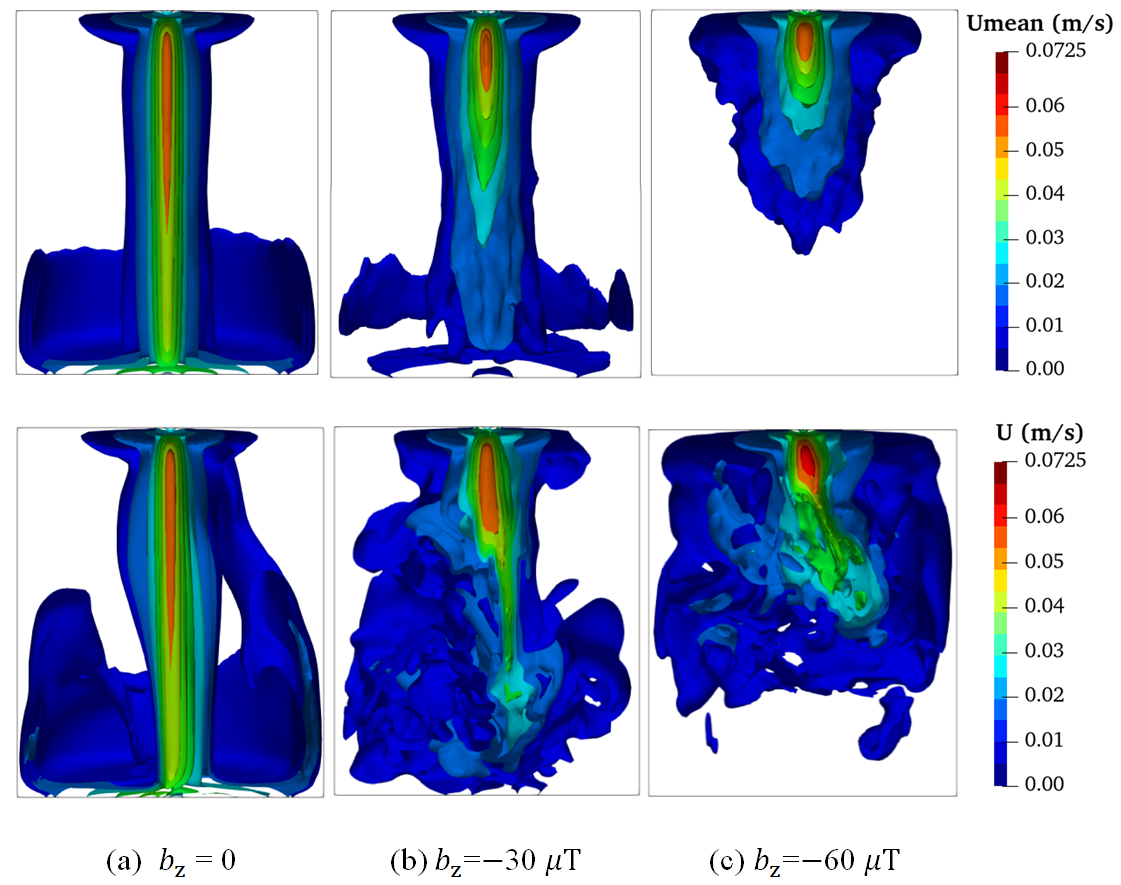}
			\caption{Cutaways of the time-averaged velocity field (upper row) and the transient velocity field (the lower row) at $t=600\hspace{1.5pt}$s for three different $b_z$.}
			\label{fig:cutaway_of_isosurfaces}
		\end{figure}

		To get a first impression of the time-averaged and instantaneous flow fields, it is convenient to use iso-surfaces of the modulus of the velocity, i.e., the flow speed (which will also be called velocity in the following). Figure \ref{fig:cutaway_of_isosurfaces} shows cutaways of the time-averaged velocity fields (upper row) and the distribution of the transient velocity field at $t=600\hspace{1.5pt}$s (lower row) for different $b_z$ configurations. Consistent with the conclusion of \cite{Ke2020}, the jet region shrinks gradually towards the top centre of the container when $b_z$ is increased. For $b_z=0$, the transient velocity field is very similar to that of the time-averaged velocity. However, drastic differences between the transient and time-averaged results are observed for $b_z \neq 0$, which means in turn that important features of the transient flow field are “smeared out” in the time-averaging procedure. In particular, transient results for $b_z \neq 0$ show much more disturbances than the time-averaged results. In addition, the jet region in the transient plots is slightly deflected from the centreline of the container, while the time-averaged results are still concentrated along the centreline, even in case of non-zero $b_z$. Hence, it is essential to take $b_z$ seriously into account even if it is much weaker than the azimuthal component caused by the current.

	  	\begin{figure}[htbp]
	  		\centering
	  		\includegraphics[width=\textwidth]{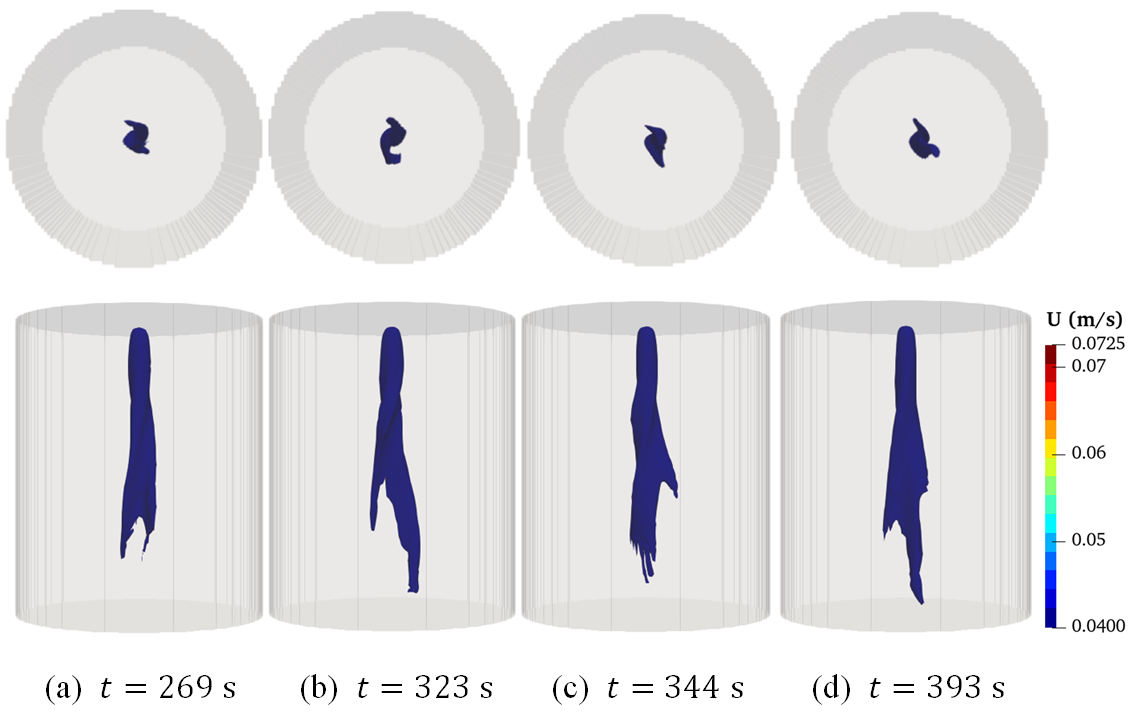}
	  		\caption{Iso-surface distributions of the higher transient velocity region ($u \geq 0.04$\hspace{1.5pt}m/s) for $b_z=-25.5\hspace{1.5pt}\mu$T at different times. The bottom views are in the upper row, and the side views are in the lower row.}
	  		\label{fig:isosurfaces_at_different_time}
	  	\end{figure}

  		\begin{figure}[htbp]
  			\centering
  			\includegraphics[width=\textwidth]{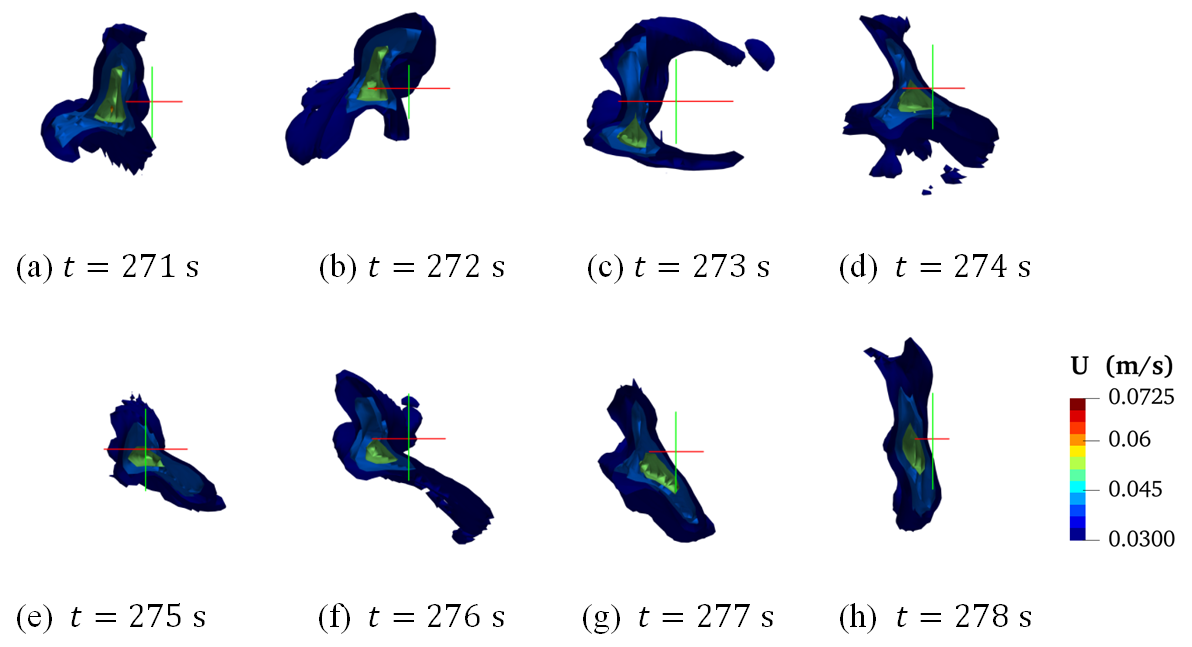}
  			\caption{Top view of transient velocity iso-surface distribution below $z=30$\hspace{1.5pt}mm at different times at $b_z=-25.5\hspace{1.5pt}\mu$T, with $u \geq 0.03$\hspace{1.5pt}m/s.}
  			\label{fig:topview_isosurfaces_at_different_time}
  		\end{figure}

		As shown in \cite{Ke2020}, when the external magnetic field is $b_z=-25.5\hspace{1.5pt}\mu$T (the other two components $b_x$ and $b_y$ have only a minor effect on the flow structure, therefore they can be ignored), the time-averaged numerical results agree well with the corresponding experimental results. In this paper, we will pay more attention to the transient behaviour at $b_z=-25.5\hspace{1.5pt}\mu$T. In order to illustrate the flow characteristics of the jet region, the iso-surfaces of the transient velocity (saved every 0.01\hspace{1.5pt}s) with the magnitude larger than 0.04\hspace{1.5pt}m/s are exported and shown in Figure \ref{fig:isosurfaces_at_different_time}. The bottom view (in positive z direction) and the side view (in positive x direction) of the fluid region at the time instants $t=$\hspace{1.5pt}269\hspace{1.5pt}s, 323\hspace{1.5pt}s, 344\,s and 393\hspace{1.5pt}s are presented in the first and second row, respectively. Although  in the bottom view the shapes of the iso-surfaces are very similar, they have quite different features from side views. More details can be observed in Figure \ref{fig:topview_isosurfaces_at_different_time}, which shows a series of snapshots from $t=$\hspace{1.5pt}271\hspace{1.5pt}s to 278\hspace{1.5pt}s. Here, the snapshots depict the top views of the transient velocity iso-surface distributions below $z=30$\hspace{1.5pt}mm, with $u \geq 0.03$\hspace{1.5pt}m/s. The tail of the jet (i.e., the high velocity region) swings irregularly around the centreline of the cylindrical container, and its shape varies from time to time. Actually, the 3D jet has a spiral structure, which will be illustrated more clearly in the streamlines presented in Figure \ref{fig:snapshots_of_velocity_streamlines} of the next Section.
	\subsection{Swirling jet flow and helicity}
\begin{figure}[htbp]
			\centering
			\includegraphics[width=\textwidth]{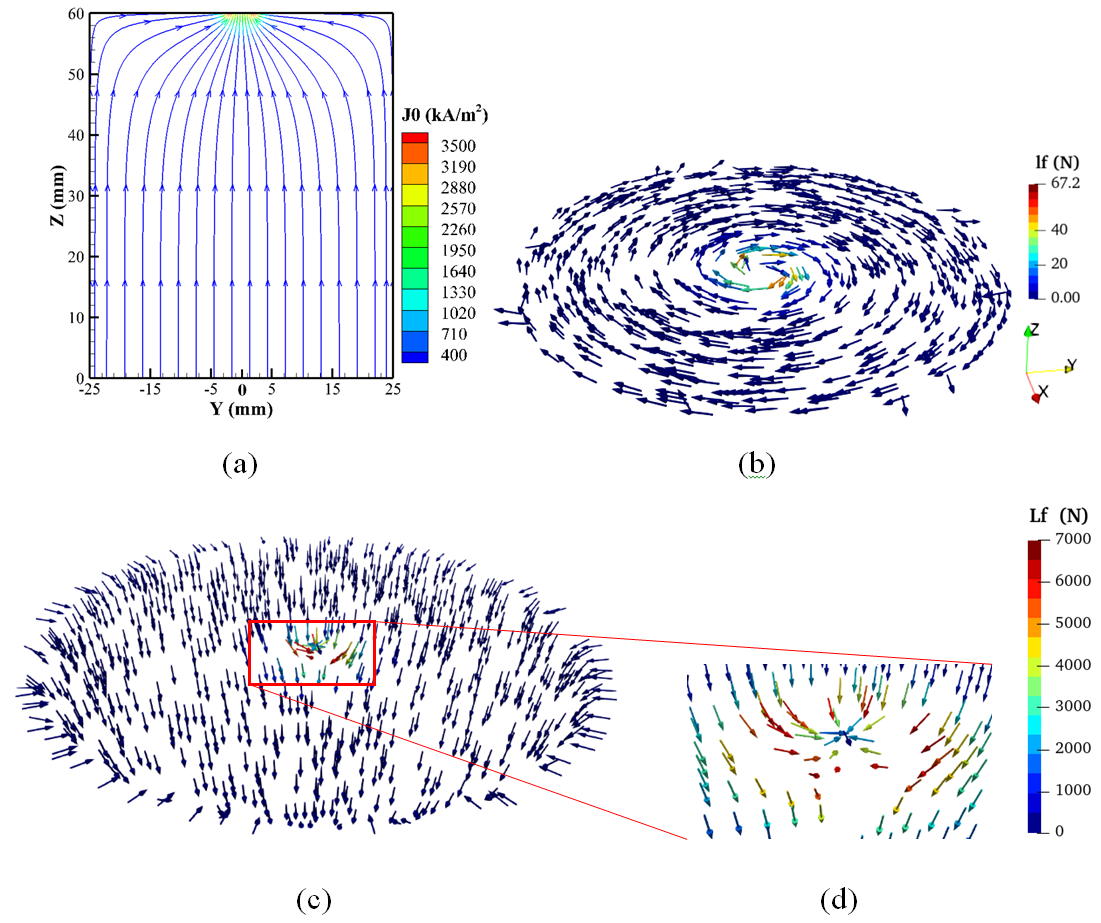}
			\caption{Distribution of (a) the imposed current density in the YOZ plane and Lorentz force on the $z=56\hspace{1.5pt}$mm cross-section for $b_z=-25.5\hspace{1.5pt}\mu$T: (b) $\bm{J}_0 \times \bm{b}_z$; (c) $\bm{J}_0 \times \bm{B}_0$; (d) partially enlarged view ($2:1$) of (c). Note the significant difference of the magnitudes between the two forces by a factor of approximately 100.}
			\label{fig:current_and_Lorentz_force}
		\end{figure}

		In order to better understand the influence of $b_z$ on the flow field structure, the distributions of the current density (due to the imposed current) in the zy-plane and of the Lorentz force in the $z=56$\,mm plane are presented in Figure \ref{fig:current_and_Lorentz_force} (for the particular value $b_z=-25.5\hspace{1.5pt}\mu$T). Due to the very different diameters of the top electrode and the cylindrical container, the current converges towards the top centre area, with a sharp rise of the magnitude of the current density $\bm{J}_0$, as shown in Figure \ref{fig:current_and_Lorentz_force}\hspace{1pt}a. Obviously, the highest current density occurs near the edge of the electrode, which will play a key role in the following analysis of the Lorentz force.

\begin{figure}[htbp]
			\centering
			\includegraphics[width=\textwidth]{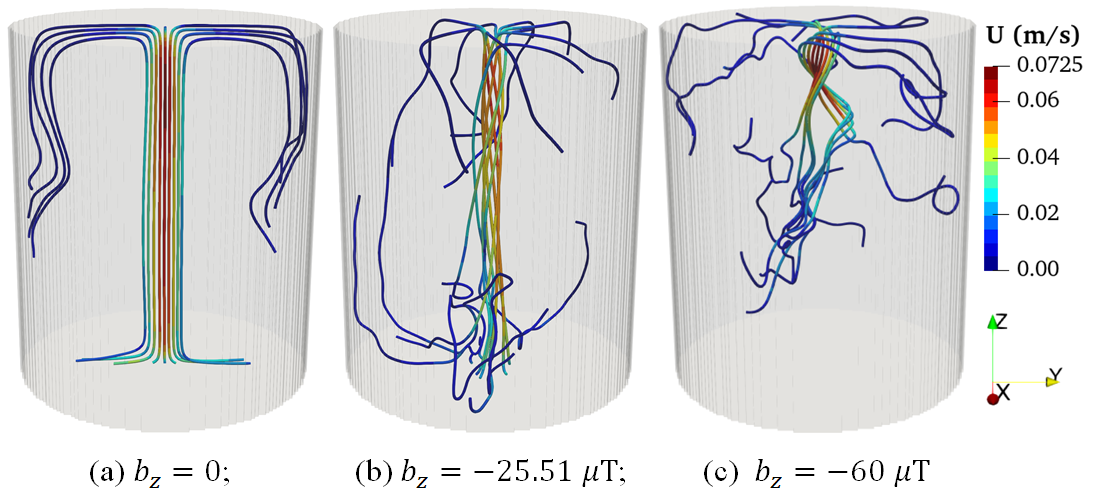}
			\caption{Snapshots of the velocity streamlines at $t=620$\hspace{1.5pt}s for different $b_z$ configuration.}
			\label{fig:snapshots_of_velocity_streamlines}
		\end{figure}

		In our numerical scheme, four parts of the Lorentz force are computed. The first and dominant part $\bm{J}_0 \times \bm{B}_0$, hereinafter referred to as \textbf{Lf}, results from the interaction of $\bm{J}_0$ with its own magnetic field $\bm{B}_0$. As shown in Figure \ref{fig:current_and_Lorentz_force}\hspace{1pt}c and d, the radially inward directed component of $\bm{J}_0$ and the toroidal $\bm{B}_0$ produce a downward-directed Lorentz force, while the z-directed component of $\bm{J}_0$ and the toroidal $\bm{B}_0$ generate a radially inward directed Lorentz force. Both together are the main sources of the jet flow, and the maximum \textbf{Lf} is generated at the same location as the maximum of $\bm{J}_0$. Second, when the liquid metal starts to flow, another part of the Lorentz force is generated (usually directed opposite to the flow) by the interaction of $\bm{B}_0$ with the current that is induced by the electromotive force $\bm{u} \times \bm{B}_0$. Driven by the first part, and modified by the second one, a jet forms close to the top centre of the container. The snapshot of the streamlines of the jet flow at $t=620$\hspace{1.5pt}s (Figure \ref{fig:snapshots_of_velocity_streamlines}\hspace{1pt}a) indicates a rather laminar and smooth flow. However, in presence of an external field $b_z$, two new parts of the Lorentz force come into play. The first new part is generated by the interaction of $\bm{J}_0$ and $\bm{b}_z$, i.e., $\bm{J}_0 \times \bm{b}_z$, and is named hereinafter \textbf{lf}. The second new part, resulting from the interaction of $\bm{b}_z$ with the current induced by the electromotive force $\bm{u} \times \bm{B}_0$, is much weaker than all other parts and will not be discussed further.

		Taking the particular case of $b_z=-25.5\hspace{1.5pt}\mu$T as an example, the distribution of \textbf{lf} on the $z=56$\hspace{1.5pt}mm plane is shown in Figure \ref{fig:current_and_Lorentz_force}\hspace{1pt}b. As discussed in \cite{Ke2020}, within the zone where $z > 51.2$\hspace{1.5pt}mm, the radial component of $\bm{J}_0$ is stronger than its axial component, hence preferably in this zone the toroidal \textbf{lf} might become strong enough to influence the flow structure. Notably, even in this zone, \textbf{Lf} is still hundreds or thousands times stronger than \textbf{lf}. Under the combined effect of radially inward directed \textbf{Lf} and the toroidal \textbf{lf}, velocity streamlines are twisted into spiral shape, as presented in Figure \ref{fig:snapshots_of_velocity_streamlines}\hspace{1pt}b. Compared with Figure \ref{fig:snapshots_of_velocity_streamlines}\hspace{1pt}a, velocity streamlines in the jet region are no longer straight and parallel, but entangled, and bifurcate in the lower part. Additionally, the lighter colour of streamlines indicates that the velocity in the lower part of the jet region is lower than that in the case of $b_z=0$. When $b_z$ reaches $-60\hspace{1.5pt} \mu$T (Figure \ref{fig:snapshots_of_velocity_streamlines}\hspace{1pt}c), velocity streamlines in the jet region are further twisted, and the jet region becomes very short. Considering both the variation of the shapes and the distributions of transient velocity streamlines with the increment of $b_z$, we deduce that the kinetic energy in the jet region is weakened with the spiral flow intensity increasing gradually.

		Comparing the numerical results shown in Figures \ref{fig:isosurfaces_at_different_time} and \ref{fig:snapshots_of_velocity_streamlines} with previous results \cite{Luginsland2012,Ivanic2003,Ianiro2011,Oberleith2011,Vanierschot2016}, with $b_z \neq 0$ the ordinary jet flow is clearly modified by the toroidal \textbf{lf} which transfers it into swirling jet flow. For example, as shown in Figure \ref{fig:snapshots_of_velocity_streamlines}, when $|b_z|$ increases from 0 to $60\hspace{1.5pt}\mu$T, the shapes of the transient velocity streamlines change in the whole fluid domain. Since \textbf{lf} decays rapidly from the top centre area to the surroundings, the jet flow starts its swirling motion already at the top centre area. Additionally, according to \cite{Li2015}, the swirling jet flow can expand the jet region outward and enhance the entrainment, which causes the attenuation of the kinetic energy. In this extreme case, the whole flow field has become highly turbulent. Consequently, the jet region is shortened towards the top centre of the container when $|b_z|$ increases, and swings around the centreline in an irregular manner.

		\begin{figure}[htbp]
			\centering
			\includegraphics[width=8.31cm,height=6.91cm]{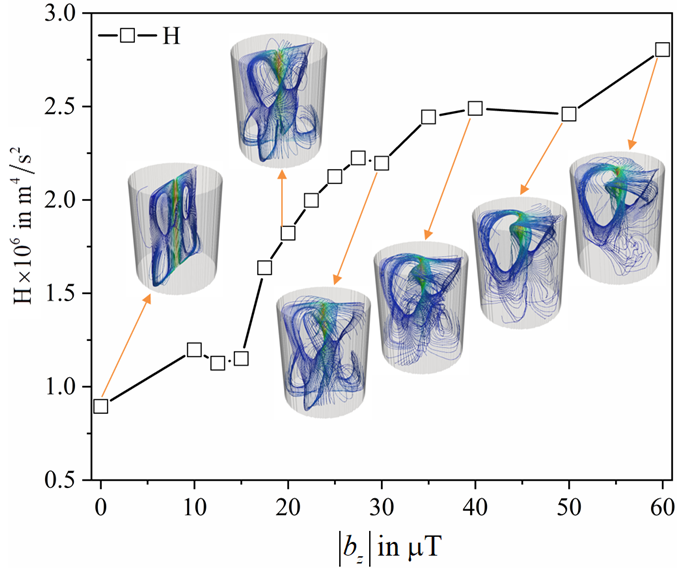}
			\caption{Helicity of transient velocity field and distributions of time-averaged velocity streamlines for different $b_z$ values.}
			\label{fig:helicities}
		\end{figure}

		In order to analyse quantitatively the swirling flow, an appropriate quantity related to the amount of velocity streamline deformation should be introduced. In fluid dynamics, the helicity is used as a measure of linkage and/or knottedness of vortex lines, as first discussed by Moffatt \cite{Moffatt1969} and Moreau \cite{Moreau1961}. In the following we will therefore evaluate the effect of $b_z$ on the helicity of the velocity field. The helicity H of a velocity field $\bm{u}$ in the volume V is defined by
		\begin{equation}
			H=\int_{V} \bm{u}\cdot(\nabla \times \bm{u}) dV.
		\end{equation}
		This quantity, in dependence on $b_z$, is shown in Figure \ref{fig:helicities}. In order to illustrate the increasingly helicoidal structure of the velocity field, the time-averaged velocity streamlines for 6 different $b_z$ are also presented in Figure \ref{fig:helicities}. As the black curve shows, H increases initially quite slowly when $|b_z|$ rises from 0 to $16\hspace{1.5pt}\mu$T, which means that the jet flow evolves into a swirling jet flow rather gradually. Then, H shows a steep increment with $|b_z|$ rising from $16\hspace{1.5pt}\mu$T to $25\hspace{1.5pt}\mu$T, after which it levels off until approximately $30\hspace{1.5pt}\mu$T. Another steep increase occurs between $30\hspace{1.5pt}\mu$T and $35\hspace{1.5pt}\mu$T, after which a sort of plateau is reached, going over to a mild increase at $60\hspace{1.5pt}\mu$T. Remarkably, this behaviour is very similar to that observed for the deviation of the velocity peak from the centreline, as shown in Figure \ref{fig:snapshots_of_velocity_streamlines}\hspace{1pt}b of \cite{Ke2020}. Hence, the helicity seems to be closely related to the radial deflection of the jet. The time-averaged velocity streamlines in Figure \ref{fig:helicities} illustrate this behaviour, showing that for increasing $|b_z|$ the streamlines become increasingly twisted. Initially, with the increase of \new{$|b_z|$} from 0 to $20\hspace{1.5pt}\mu$T, the shapes of velocity streamlines change from a 2D to a 3D structure, but they almost keep their features when $|b_z|$ is stronger than $50\hspace{1.5pt}\mu$T. It seems that for those stronger fields the additional kinetic energy generated by \textbf{lf} is just balanced or neutralized by the increased dissipation due to the additional swirling motion. As discussed in \cite{Ke2020}, the measured $b_z$ is about $-25.5\hspace{1.5pt}\mu$T (although the exact value is not easily determinable due to presence of magnetic screws close to the upper lid \cite{Ke2020,Starace2015}). It can be seen from Figure \ref{fig:helicities} that this value is just in the range where there is the strongest impact on the flow field.

\begin{figure}[htbp]
			\centering
			\includegraphics[width=8.29cm,height=6.86cm]{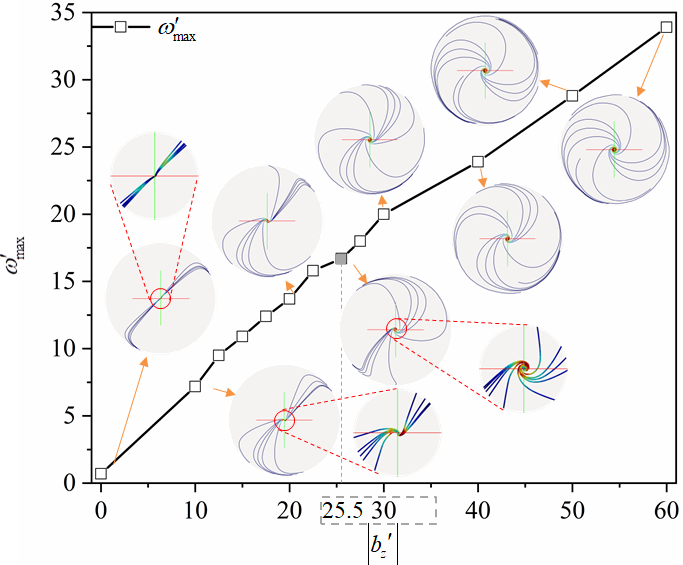}
			\caption{The maximum of the time-averaged angular velocities and the streamline distributions in the $z=58$\hspace{1.5pt}mm cross section for different $b_z$ values.}
			\label{fig:angular_velocity}
		\end{figure}

		The typical swirling motion in the top area is illustrated in Figure \ref{fig:angular_velocity} which shows the 2D time-averaged velocity streamlines and maximum angular velocities in the $z=58\hspace{1.5pt}$mm plane. Without any $b_z$, the streamlines are almost straight and directed towards the centre, except the backflow area near the side wall. In the whole plane, the streamlines become gradually curly with increasing $|b_z|$. When $|b_z|$ is larger than $30\hspace{1.5pt}\mu$T, all streamlines bend in a clockwise direction, which looks like a windmill whose rotation axis is the centre of the circular cross section. From the partially enlarged view, the streamlines roll up more intensively in the centre area with the increment of $|b_z|$. In order to quantify the rotating rate, the angular velocity, $\bm{\omega}$, is computed by
		\begin{equation}
			\bm{\omega}=\frac{\bm{r} \times \bm{u}_{xy}}{r^2},
		\end{equation}
		where $\bm{r}$ is the position vector in the points in the $z=58\hspace{1.5pt}$mm plane measured from the centre, and $\bm{u}_{xy}$ is the horizontal velocity. Possessing a similar distribution as the toroidal Lorentz force \textbf{lf}, the maximum angular velocity (in red) usually occurs close to the centre area, where the jet flow starts to rush downwards. In other words, this area is the origin of the swirling jet flow.

		In a first attempt, if the maximum angular velocity $\bm{\omega}_{\rm max}$ and $\bm{b}_z$ are nondimensionalised by 1\hspace{1.5pt}rad/s and $1\hspace{1.5pt}\mu$T as $\bm{\omega}_{\rm max}^\prime$ and $\bm{b}_z^\prime$ respectively, then a relationship between their magnitudes, $|\omega_{\rm max}^\prime|$ and $|b_z^\prime|$, can be approximated by a piece-wise linear function according to
		\begin{equation}
			|\omega_{\rm max}^\prime|=\left\{
			\begin{array}{rcl}
				0.64|b_z^\prime|+0.7, && {|b_z^\prime| \leq 30} \\
				0.50|b_z^\prime|+3.9, && 30 < {|b_z^\prime| \leq 60}
			\end{array}\right.
		\end{equation}

		\begin{figure}[htbp]
			\centering
			\includegraphics[width=\textwidth]{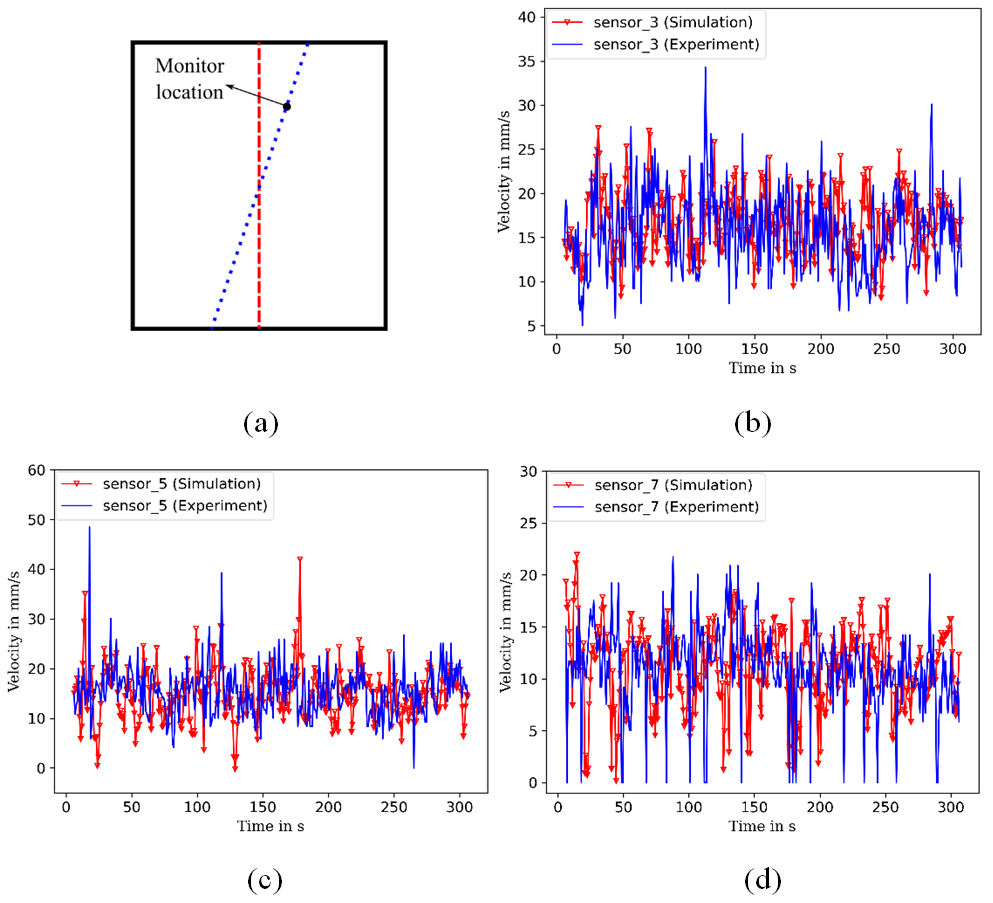}
			\caption{Comparisons of the time evolutions of the velocity on different UDV beam lines between numerical and experimental results (for $b_z=-25.5\hspace{1.5pt}\mu$T): (a) Schematic diagram of the monitor location; (b) sensor 3; (c) sensor 5; (d) sensor 7.}
			\label{fig:comparisons}
		\end{figure} 
		
	\subsection{Comparisons of transient velocity fields}
		In the following, the temporal evolution of the numerically determined velocity is compared with the experimental data at some particular monitor locations on the UDV beam lines for the particular value $b_z=-25.5\hspace{1.5pt}\mu$T (Figure \ref{fig:comparisons}). We observe, first, that the velocity oscillates quite irregularly. Second, the numerical results are in good agreement with the measured data, both with respect to the average value and to the fluctuation level. Third, numerical and experimental velocities, obtained by sensor 3 and 5, are a little higher than those measured by sensor 7. This indicates that the jet (and the high velocity region related to it) moves preferably to one side of the cylindrical container. The reason for that might be either connected with the horizontal components of the Earth’s magnetic field or by slight asymmetries of the experimental setting.

	\section{Conclusions and Outlook}
		In this paper, the transient behaviour of the Electro-Vortex flow (EVF) in a cylindrical container under the influence of an external magnetic field ($b_z$) was investigated by numerical simulation and experiment, following our former experimental and time-averaged numerical study in \cite{Ke2020}. Our main conclusions are as follows:

		First, in accordance with \cite{Ke2020}, the jet region (with high velocity) becomes shorter with increasing $b_z$. The transient velocity distributions reveal now that the surrounding fluid is not at rest, but is stirred by the swirling jet when $b_z \neq 0$. The motion of the jet around the centreline of the cylinder is not regular, but varies non-periodically, with the tail of the high velocity region showing bifurcations with different shapes and locations.

		Second, the radially inward directed and downward directed Lorentz force \textbf{Lf} drives the liquid metal into a jet flow, while the toroidal Lorentz force \textbf{lf} rotates the jet into a swirling motion. Although $b_z$ is very weak, it has significant effects. When $|b_z|$ grows from 0 to $60\hspace{1.5pt}\mu$T, the jet spirals more intensely and shrinks gradually towards the top centre of the container. The spiral intensity and the rotation frequency are evaluated by the helicity magnitude and the maximum angular velocity, respectively.

		At last, numerical transient velocities at fixed points are compared with the experimental data, which shows generally a good agreement. In the future, we may also study the extreme regime where the poloidal flow structure is nearly completely suppressed by the toroidal motion, as recently shown in \cite{Kolesnich2020}.

	\section*{Acknowledgments}
	This work was supported by Deutsche Forschungsgemeinschaft (DFG, German Research Foundation) under award No. 338560565, in frame of the Helmholtz–RSF Joint Research Group “Magnetohydrodynamic Instabilities: Crucial relevance for large scale liquid metal batteries and the sun-climate connection”, contract No. HRSF-0044 and RSF-18-41-06201, by National Natural Science Foundation of China, No. 51976021, and by The Fundamental Research Funds for the Central Universities, No. DUT21GJ202.
	
	\newcommand{\noopsort}[1]{}

	\lastpageno
\end{document}